\documentclass[showpacs, aps, prb, unsortedaddress]{revtex4-1}

\usepackage{amssymb}
\usepackage{latexsym}
\usepackage{amsmath}
\usepackage{graphicx}
\usepackage{color}
\begin{document}

\title{Level Repulsion and Band Sorting in Phononic Crystals}

\date{\today}
\author{Yan Lu, Ankit Srivastava}
\affiliation{Department of Mechanical, Materials, and Aerospace Engineering,
Illinois Institute of Technology, Chicago, IL, 60616
USA}
\email{asriva13@iit.edu}

\begin{abstract}
In this paper we consider the problem of avoided crossings (level repulsion) in phononic crystals and suggest a computationally efficient strategy to distinguish them from normal cross points. This process is essential for the correct sorting of the phononic bands and, subsequently, for the accurate determination of mode continuation, group velocities, and emergent properties which depend on them such as thermal conductivity. Through explicit phononic calculations using generalized Rayleigh quotient, we identify exact locations of exceptional points in the complex wavenumber domain which results in level repulsion in the real domain. We show that in the vicinity of the exceptional point the relevant phononic eigenvalue surfaces resemble the surfaces of a 2 by 2 parameter-dependent matrix. Along a closed loop encircling the exceptional point we show that the phononic eigenvalues are exchanged, just as they are for the 2 by 2 matrix case. However, the behavior of the associated eigenvectors is shown to be more complex in the phononic case. Along a closed loop around an exceptional point, we show that the eigenvectors can flip signs multiple times unlike a 2 by 2 matrix where the flip of sign occurs only once. Finally, we exploit these eigenvector sign flips around exceptional points to propose a simple and efficient method of distinguishing them from normal crosses and of correctly sorting the band-structure. Our proposed method is roughly an order-of magnitude faster than the zoom-in method and correctly identifies $>97\%$ of the cases considered. Both its speed and accuracy can be further improved and we suggest some ways of achieving this. Our method is general and, as such, would be directly applicable to other eigenvalue problems where the eigenspectrum needs to be correctly sorted.

\end{abstract}

\maketitle

\section{Introduction}

Phononic crystals are artificially designed periodic composite materials \cite{pennec2010two,lee2012micro,hussein2014dynamics,srivastava2015elastic} which can significantly affect the propagation characteristics of acoustic and stress waves through them in 1- \cite{nemat2015anti,willis2015negative,srivastava2016metamaterial,srivastava2017evanescent,shmuel2016universality,chen2016modulating}, 2- \cite{shim2015harnessing,mousanezhad2015honeycomb}, and 3-dimensions \cite{babaee2015three,galich2016shear,lu20173}. Although the behavior of phononic crystals can be analyzed through full Finite Element simulations, a vitally important tool in phononics research is band-structure calculation. In its very basic form, the phononic band-structure is the dispersion relation of the phononic crystal derived by calculating the real frequency eigenvalues ($\omega$) of the equation of motion (acoustic or elastodynamic) for given values of real wavevector ($\mathbf{k}$) which span the boundaries of the crystal's Irreducible Brillioun Zone (IBZ). It should be noted, however, that more complex bandstructures with complex values of frequency and/or wavevector are possible as well\cite{trainiti2016non,frazier2016generalized,mazzotti2017band,krattiger2016anisotropic}. The (real $\omega$-real $\mathbf{k}$) band-structure (phononic or photonic) represents a parametric and Hermitian eigenvalue problem where the frequency eigenvalue depends parametrically upon the wavevector. Any parametric eigenvalue problem exhibits two phenomena that are of interest to us \cite{kato1966perturbation,heiss1990avoided,heiss1991transitional}:
\begin{itemize}
\item Diabolic points: These are values of the parameter in the real domain which exhibit 2 degenerate eigenvalues whose associated eigenvectors are, nevertheless, linearly independent.
\item Exceptional points: These are values of the parameter in the complex domain of the parameter which exhibit degenerate eigenvalues as well as degenerate associated eigenvectors.
\end{itemize}
It is well known that eigenvalues are repulsed (level repulsion\cite{wu2004level,yeh2016level}) from each other in the vicinity of exceptional points \cite{heiss1990avoided}. Therefore, the existence of exceptional points in the vicinity of the real domain of the parameter will cause the associated eigenvalues to avoid each other in the real domain. Furthermore, in the real domain the eigenvectors of the associated eigenvalues will be rapidly exchanged in the level repulsion region. This rapid exchange of eigenvectors in the level repulsion region has been used to practical effect in the design of highly sensitive sensors in discrete and lower dimensional systems\cite{manav2014ultrasensitive,foreman2013level}. In the context of phononics and photonics, diabolic points correspond to real crosses of eigenvalue branches in the band-structure and exceptional points are associated with avoided crossings of the branches\cite{ding2016emergence,lin2016enhanced}. Branch continuations in the vicinity of crosses (real or apparent) become difficult to determine with coarse wavevector discretizations and it vitally affects some application areas of phononics. One area of such relevance is in employing phononic crystals for designing materials with extreme thermal properties\cite{davis2014nanophononic,xiong2016blocking}. It has been shown that the correct sorting of the phononic band-structure is of prime importance in the accurate prediction of thermal properties of phononic crystals\cite{zen2014engineering,anufriev2015thermal}. Since thermal properties directly depend upon the group velocities of phononic bands, incorrect band sorting can lead to more than 40\% error in the predicted values of phonon emission power calculated over a coarse wave-vector grid\cite{zen2014engineering}. 

The techniques which have been applied for phononics band-sorting in literature are essentially based upon eigenvector orthogonality\cite{huang2014correlation,zen2014engineering} and polarizations of modes\cite{wu2004level,achaoui2010polarization}. However, since eigenvectors and polarizations are exchanged in level repulsion regions, all such techniques are bound to fail in distinguishing real crosses from avoided crosses as we clearly show in this paper. The only fail-safe method proposed in literature to make this distinction is to calculate the eigenvalues of the system over very fine parametric discretizations in the crossing zones, thereby, explicitly distinguishing real crosses from apparent ones\cite{wu2004level,achaoui2010polarization}. This process is highly computationally inefficient and simply infeasible in cases where thousands of such cross points might need to be considered, as is the case for thermal properties. 

In this paper, we propose a new and highly computationally efficient method of distinguishing real crosses from avoided ones for the phononic bandstructure. The technique can be extended to any other system where eigenvalue bands need to be correctly sorted. Our method is based on the principal that eigenvectors suffer flips of signs along closed contours around exceptional points\cite{heiss1999phases,heiss2000repulsion}. A large enough closed curve near a cross will either contain an exceptional point (if it exists) or not (if it doesn't exist). Therefore, by calculating the eigenvectors at some points on this closed contour we can distinguish real crosses from apparent ones with near certainty. The method does not require us to actually find the location of the exceptional point as it only depends upon making the closed contour large enough to encircle it. 

In the subsequent sections we first recast the phononic problem in a self-adjoint form ensuring that its associated finite dimensional representation will be Hermitian for real values of the wavevector. We then use the generalized Rayleigh quotient to explicitly identify an exceptional point in the complex wavevector domain. We show that in the vicinity of the exceptional point, the behavior of the eigenvalues resembles that of a $2\times 2$ matrix. As expected, the eigenvalues are exchanged along a closed curve around it. However, unlike for a $2\times 2$ matrix, for some exceptional points we show that the eigenvectors flip signs multiple times. 
%The number of flips is still consistent with the prevailing knowledge that along a closed curve encircling an exceptional point, one of the eigenvectors flips sign and the other does not. 
We show that no such flips occur when the closed curve either does not encircle an exceptional point or pass a branch cut. Finally, we propose our method for distinguishing real crosses from apparent ones and use it to correctly sort an extended phononic band-structure.

\section{Problem Description}

\begin{figure}[htp]
\centering
\includegraphics[scale=1]{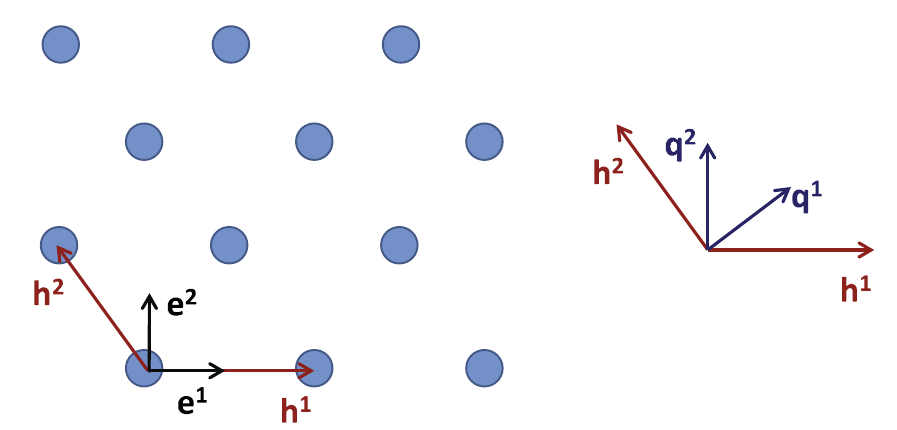}
\caption{Schematic of a 2-dimensional periodic composite. The unit cell vectors ($\mathbf{h}^1,\mathbf{h}^2$), reciprocal basis vectors ($\mathbf{q}^1,\mathbf{q}^2$), and the orthogonal vectors ($\mathbf{e}^1,\mathbf{e}^2$) are shown.}\label{fVectors}
\end{figure}

The frequency domain dynamics of a linear elastic medium with a spatially varying constitutive tensor $\mathbf{C}$ and density $\rho$ is given by:
\begin{equation}
\boldsymbol{\Lambda}(\mathbf{u})+\mathbf{f}=\lambda\mathbf{u};\quad \boldsymbol{\Lambda}(\mathbf{u})\equiv \frac{1}{\rho}\left[C_{ijkl}u_{k,l}\right]_{,j}
\end{equation}
where $\lambda=-\omega^2$, $\mathbf{u}$ is the displacement field, $\mathbf{f}$ is the body force, and $\boldsymbol{\Lambda}$ is a linear differential operator. The operator operates on the Hilbert space of square integrable vector functions over which the inner product is defined with respect to the weight function $\rho$:
\begin{equation}
\langle\mathbf{v},\mathbf{w}\rangle=\int_\Omega \rho\mathbf{v}^*\cdot \mathbf{w}\mathrm{d}\Omega
\end{equation}
It can be shown that the operator is self adjoint ($\langle\mathbf{v},\boldsymbol{\Lambda}(\mathbf{w})\rangle=\langle\boldsymbol{\Lambda}(\mathbf{v}),\mathbf{w}\rangle$). For phononic problems the problem domain is periodic and is defined by a repeating unit cell. For the general 3-dimensional case, the unit cell ($\Omega$) is characterized by 3 base vectors $\mathbf{h}^i$, $i=1,2,3$. Any point within the unit cell can be uniquely specified by the vector $\mathbf{x}=H_i\mathbf{h}^i=x_i\mathbf{e}^i$ where $\mathbf{e}^i$ are the orthogonal unit vectors and $0\leq H_i\leq 1,\forall i$. The unit cell is associated with a set of reciprocal base vectors $\mathbf{q}^i$ such that $\mathbf{q}^i\cdot\mathbf{h}^j=2\pi\delta_{ij}$. Reciprocal lattice vectors are represented as a linear combination of the reciprocal base vectors, $\mathbf{G}^\mathbf{n}=n_i\mathbf{q}^i$, where $n_i$ are integers. Fig. (\ref{fVectors}) shows the schematic of a 2-D unit cell, clearly indicating the unit cell basis vectors, the reciprocal basis vectors and the orthogonal basis vectors.
The material properties have the following periodicity:
\begin{equation}
C_{jkmn}(\mathbf{x}+n_i\mathbf{h}^i)=C_{jkmn}(\mathbf{x});\quad \rho(\mathbf{x}+n_i\mathbf{h}^i)=\rho(\mathbf{x})
\end{equation}
where $n_i(i=1,2,3)$ are integers. Due to the periodic nature of the problem, it accepts solutions of the form $\mathbf{u}(\mathbf{x})=\mathbf{u}^p(\mathbf{x})e^{\mathrm{i}\mathbf{k}.\mathbf{x}}$ where $\mathbf{k}$ is the Bloch wavevector with components $\mathbf{k}=Q_i\mathbf{q}^i$ where $0\leq Q_i\leq 1,\forall i$ and $\mathbf{u}^p$ is a periodic function. Under the substitution, the harmonic elastodynamic problem can be formally written as (neglecting the body force):
\begin{equation}
\boldsymbol{\Lambda}^{(\mathbf{k})}(\mathbf{u}^p)=\lambda\mathbf{u}^p
\end{equation}
where the superscript $(\mathbf{k})$ is now included to emphasize that the operator depends upon the Bloch wavevector. Explicitly we have:
\begin{equation}
\boldsymbol{\Lambda}^{(\mathbf{k})}(\mathbf{u}^p)\equiv \frac{1}{\rho}\left[C_{ijkl}u_{k,l}^p\right]_{,j}+\frac{\mathrm{i}q_jC_{ijkl}}{\rho}u_{k,l}^p+\frac{\mathrm{i}q_l}{\rho}\left[C_{ijkl}u_k^p\right]_{,j}-\frac{q_lq_jC_{ijkl}}{\rho}u_k^p
\end{equation}
It can be shown that for real values of the wavevector the operator is self adjoint:
\begin{equation}
\langle\mathbf{v}^p,\boldsymbol{\Lambda}(\mathbf{w}^p)\rangle=\langle\boldsymbol{\Lambda}(\mathbf{v}^p),\mathbf{w}^p\rangle
\end{equation}
It, therefore, has real eigenvalues $\lambda_n$ for real values of the wavevector. Furthermore, its eigenfunctions , $\tilde{\mathbf{u}}_n^p$, form a complete and orthonormal set:
\begin{eqnarray}
\displaystyle \nonumber \langle u_{n;i}^p(\mathbf{x}),(u_{n;j}^p)^*(\mathbf{x}')\rangle=\delta_{ij}\delta(\mathbf{x}-\mathbf{x}')\\
\displaystyle \langle\mathbf{u}_m^p,\mathbf{u}_n^p\rangle=\delta_{mn}
\end{eqnarray}
For a suitable span of the wavevector, the sets of eigenvalues $\lambda_n$ (and the corresponding frequencies $\omega_n$) constitute the phononic dispersion relation of the composite. There are several numerical techniques for calculating the eigenvalues but a common method is to expand the field variable $\mathbf{u}^p$ in an appropriate basis and then use the basis to convert the differential equation into a set of linear equations. Both the Plane Wave Expansion\cite{kushwaha1993acoustic} method and Rayleigh quotient\cite{lu2016variational} follow this strategy. As an example, trigonometric basis can be used to write $\mathbf{u}^p=\mathbf{U}^\mathbf{n}e^{\mathrm{i}\mathbf{G^n}\cdot\mathbf{x}}$ where Einstein summation is implied. $\mathbf{U}^\mathbf{n}$ is a constant vector with components $U^\mathbf{n}_i;\; i=1,2,3$. Each component depends upon the appropriate Fourier component which, in 3-dimension, refers to the set $(n_1n_2n_3)$. Either through a variational law or by expanding materials properties in the same basis, we arrive at a system of linear equations formally written as:
\begin{equation}
\left[\boldsymbol{\Lambda}^{(\mathbf{k})}\right]\{\mathbf{U}^\mathbf{n}\}=\lambda\{\mathbf{U}^\mathbf{n}\}
\end{equation}
which is a simple matrix eigenvalue problem. The $a^\mathrm{th}$ eigenvalue $\lambda_a$ has the associated eigenvector $\mathbf{U}^{\mathbf{n};a}$ which provides the $a^\mathrm{th}$ eigenfunction of the original problem through $\mathbf{u}^p_a=\mathbf{U}^{\mathbf{n};a}e^{\mathrm{i}\mathbf{G^n}\cdot\mathbf{x}}$. The orthogonality condition now reads:
\begin{eqnarray}
\displaystyle \langle\mathbf{u}_a^p,\mathbf{u}_b^p\rangle=\langle U^{\mathbf{m};a}_ie^{\mathrm{i}\mathbf{G^m}\cdot\mathbf{x}},U^{\mathbf{n};b}_ie^{\mathrm{i}\mathbf{G^n}\cdot\mathbf{x}}\rangle=\delta_{ab}
\end{eqnarray}
The above can be simplified further by noting that the basis functions themselves are orthogonal in the following sense: 
\begin{equation}
\int_\Omega e^{\mathrm{i}[\mathbf{G^m}-\mathbf{G^n}]\cdot \mathbf{x}}d\Omega=V\int_0^1\int_0^1\int_0^1 e^{\mathrm{i}2\pi[\alpha H_1+\beta H_2+\gamma H_3]}dH_1dH_2dH_3=V\delta_{\mathbf{m}\mathbf{n}}
\end{equation}
where $\alpha,\beta,\gamma$ are integers and $V$ is the volume of the unit cell. Employing this we have the following expression for the orthogonality of the eigenvectors:
\begin{eqnarray}
\displaystyle \langle U^{\mathbf{m};a}_i,U^{\mathbf{m};b}_i\rangle=\frac{\delta_{ab}}{V}
\end{eqnarray}
For the rest of the paper we have normalized the eigenvectors such that 
\begin{eqnarray}\label{eOrtho}
\displaystyle \langle U^{\mathbf{m};a}_i,U^{\mathbf{m};b}_i\rangle=\delta_{ab}
\end{eqnarray}

\subsection{Example}

\begin{figure}[htp]
\centering
\includegraphics[scale=0.6]{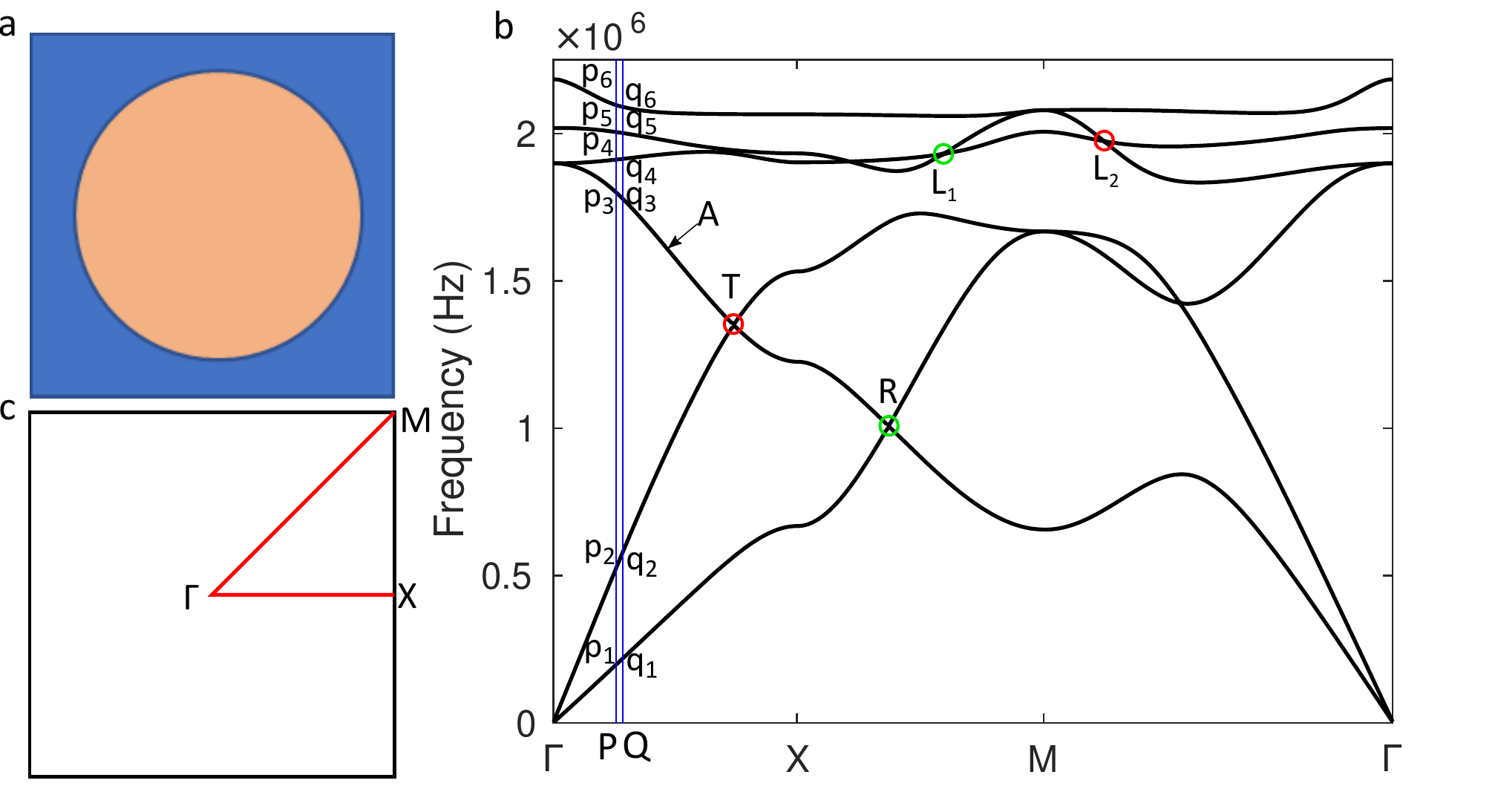}
\caption{(a) Square unit cell consists of PMMA (orange) and Nickel (blue) regions with the corresponding first Brillouin zone shown in (c). The band-structure is shown in (b) and is calculated along the boundaries of IBZ. The eigenvectors belonging to wavevector points P ($Q_1=0.1291$, $Q_2=0$) and Q ($Q_1=0.1426$, $Q_2=0$) are used in the inner product calculations.}\label{unitcell}
\end{figure}
As an example \cite{wu2004level}, a 2-D phononic composite is considered whose unit cell is shown in Fig. (\ref{unitcell}a). The unit cell is composed of Nickle ($C_{11}=324\text{ GPa},C44=80\text{ GPa},\rho=8905\text{kg/m}^3$) matrix with embedded circular PMMA ($C_{11}=7.11\text{ GPa},C44=2.03\text{ GPa},\rho=1200\text{kg/m}^3$) inclusions in a square lattice. The lattice constant is $1mm$ with an inclusion fill fraction of $50\%$. Fig. (\ref{unitcell}b) depicts the corresponding phononic dispersion relation for in-plane modes along the boundary of the Irreducible Brillouin Zone (IBZ) (Fig. \ref{unitcell}c). This band structure, which is rendered from the self-adjoint form of elastodynamic eigenvalue problem using Rayleigh quotient, shows very good agreement with previously published plane wave expansion (PWE) results \cite{wu2004level}.
The figure shows two vertical lines which denote adjacent wavenumbers over which the eigenvalue computations have been performed. Within the frequency range under consideration, the eigenvalue solutions at these wavenumbers include 6 solution points ($p_i,q_i;\;i=1,...6$). The corresponding eigenvectors obviously satisfy Eq. (\ref{eOrtho}) when they all belong to the same wavenumber. However, if the wavenumbers P and Q are close enough to each other then it is expected that Eq. (\ref{eOrtho}) will still be approximately satisfied. In other words, if $\mathbf{U}^{\mathbf{m};a}$ belongs to P and $\mathbf{U}^{\mathbf{m};b}$ belongs to Q then we should still have:
\begin{eqnarray}\label{eOrthoApp}
\displaystyle \sum_{i=1}^3\left[U^{\mathbf{m};a}_i\right]^*U^{\mathbf{m};b}_i\approx\delta_{ab}
\end{eqnarray}
if $P,Q$ are close to each other. This approximate orthogonality of the eigenvectors of two neighboring wavenumbers can be used to sort the modes of the phononic bandstructure in regions which are away from the regions of level repulsion. The logic behind this is that the modes which form a single phononic band over the span of the wavevector will have associated eigenvectors which produce nearly unitary inner products with their neighboring modes on the same band. They will, on the other hand, produce nearly zero values when their inner products are taken with neighboring modes on other bands. As an example we present the calculated absolute values of the inner products between modes $p_{1-6}$ and $q_{1-6}$ in Table \ref{innerproducts}. The table confirms that approximate orthogonality holds for the considered modes.
\begin{table}[htp]
\caption{The absolute values of inner products of $p_{1-6}$ and $q_{1-6}$.}\label{innerproducts}
\centering
\begin{tabular}{lcccccc}
\hline & $q_1$ & $q_2$ & $q_3$ & $q_4$ & $q_5$ & $q_6$\\
\hline $p_1$ & 0.9999 & 4.8268e-12 & 0.0033 & 1.6292e-13 & 0.0012 & 3.4603e-5\\
$p_2$ & 4.4115e-12	& 1.0000	& 5.3649e-13 &	0.0033 & 5.0022e-14 & 7.8003e-14\\
$p_3$ & 0.0033 & 4.4004e-13 & 0.9985 & 9.9670e-13 & 0.0453	&0.0282\\
$p_4$ & 1.7215e-13 & 0.0034 & 5.0500e-13 & 0.9999 &	3.7333e-13 & 9.8231e-13\\
$p_5$& 9.6713e-4 & 1.0209e-13 & 0.0432 & 2.5041e-13	& 0.9965 & 0.0676\\
$p_6$ & 2.3906e-6 & 5.1212e-14 & 0.0311 & 9.7758e-13 & 0.0662 & 0.9972\\
\hline
\end{tabular}
\end{table}
After sorting using this approximate orthogonality, the dispersion curves appear to cross each other at ten points. \emph{However, not all of these cross points are true cross points}.   
\begin{figure}[htp]
\centering
\includegraphics[scale=.45]{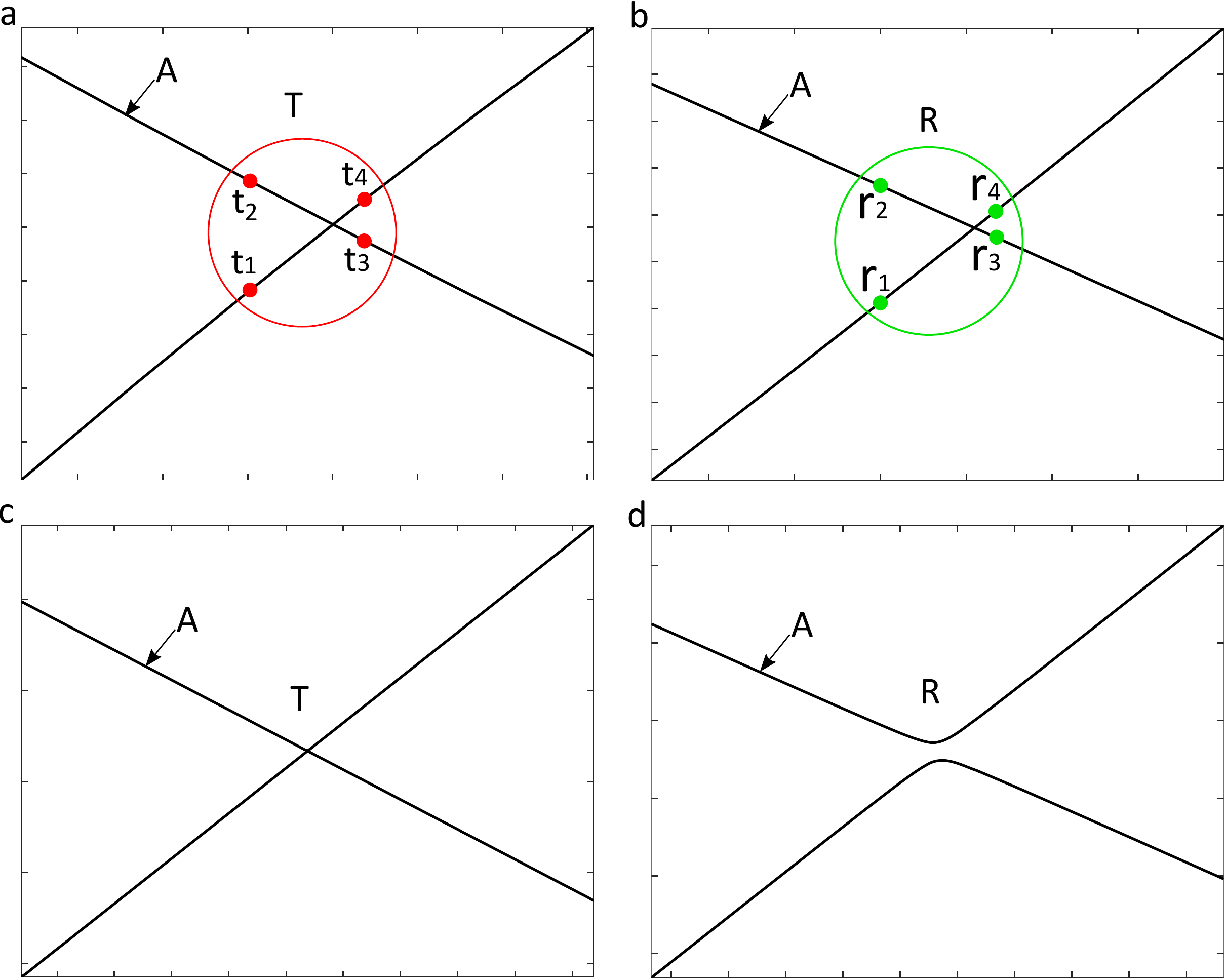}
\caption{(a),(b) Zoom-in to Point T and R in Fig. (\ref{unitcell}b). $t_i$ and $r_i$ are neighboring modes on coarse discretization of wavenumbers. $t_1$ and $t_2$ belong to the same wavenumber $Q_1=0.3651, Q_2=0$, $t_3$ and $t_4$ belong to $Q_1=0.3719, Q_2=0$, $r_1$ and $r_2$ belong to $Q_1=0.5, Q_2=0.18$, $r_3$ and $r_4$ belong to $Q_1=0.5, Q_2=0.1867$. (c),(d) Band structure evaluations on finer discretization of wavenumbers. }\label{R1T1}
\end{figure}

To clarify this further we focus on the two points marked by T and R in Fig. (\ref{unitcell}) (and Figs. \ref{R1T1}a,b). The modes around these cross points are marked as $t_i$ and $r_i$ respectively, where $i=1,\cdots,4$. $t_1, t_2$ and $t_3, t_4$ belong to two different wavevectors which are separated from each other by a distance of $6.6667\times10^{-3}$ in the reduced wavenumber domain. The same situation applies to $r_i$. The inner products between $t_1,t_4$ and $t_2,t_3$ are both 0.9999. Between $t_1,t_3$ and $t_2,t_4$ they are close to zero. Similarly, the inner products between $r_1,r_4$ and $r_2,r_3$ are both 1.0000 and between $t_1,t_3$ and $t_2,t_4$ they are close to zero. The inner products indicate that at this level of wavevector discretization, the two points should be determined as real cross points. Figs. (\ref{R1T1}c,d) show zoomed-in views of the T and R points with the wavenumber regions shown in Figs. (\ref{R1T1}a,b) further discretized in about 1000 intermediate points. The zoomed-in views clearly show that while at T the two curves indeed cross each other, at R the curves move apart. R is an apparent cross point when the IBZ is roughly discretized but is, in reality, an avoided crossing. There are several other apparent cross points in the bandstructure in Fig. (\ref{unitcell}b). These are marked off with green circles whereas the real cross points are marked off with red circles. The apparent cross points can always be distinguished from the real cross points by calculating the bandstructure over a very fine grid of wavevectors. This has been the dominant approach in literature\cite{wu2004level,achaoui2010polarization}. However, this approach is computationally wasteful and we would like to be able to distinguish apparent cross points from real cross points for accurate band sorting over a rough grid of wavevectors. 

\section{Level Repulsion in Phononic Crystals}

Consider an $N\times N$ matrix $\mathbf{A}(\mu)$ which depends upon a parameter $\mu$. For real values of $\mu$ consider that this matrix is self-adjoint. It, therefore, has $N$ real eigenvalues $\lambda_n,\;n=1,...N$ which are determined by the solution of the equation $\mathrm{det}(\mathbf{A}(\mu)-\lambda\mathbf{I})=0$ for real $\mu$. Now extend the domain of $\mu$ to the complex plane. In this case, $\mathbf{A}$ generally ceases to be self-adjoint and the eigenvalues become complex. In fact, it can be shown that the solution of $\mathrm{det}(\mathbf{A}(\mu)-\lambda\mathbf{I})=0$ is a multi-valued function $\lambda(\mu)$ which is complex valued and is defined over the complex plane\cite{heiss1990avoided,heiss1991transitional}. It has a Riemannian sheet structure with $N$ Riemann sheets which are connected with each other through square root branch points \cite{heiss1990avoided,heiss1991transitional}. The branch points are termed \emph{exceptional points}. At these points we have the coalescing of both a set of eigenvalues and their associated eigenvectors. This is in contrast with such points as the cross point T in Fig. (\ref{R1T1}) where there is a degeneracy of two eigenvalues but the associated eigenvectors are still linearly independent. The latter is called a \emph{diabolic point}. The presence of exceptional points in the complex $\mu$ plane is associated with avoided crossings (level repulsion) of $\lambda_n$ on the real $\mu$ axis. 

\begin{figure}[htp]
\centering
\includegraphics[scale=.5]{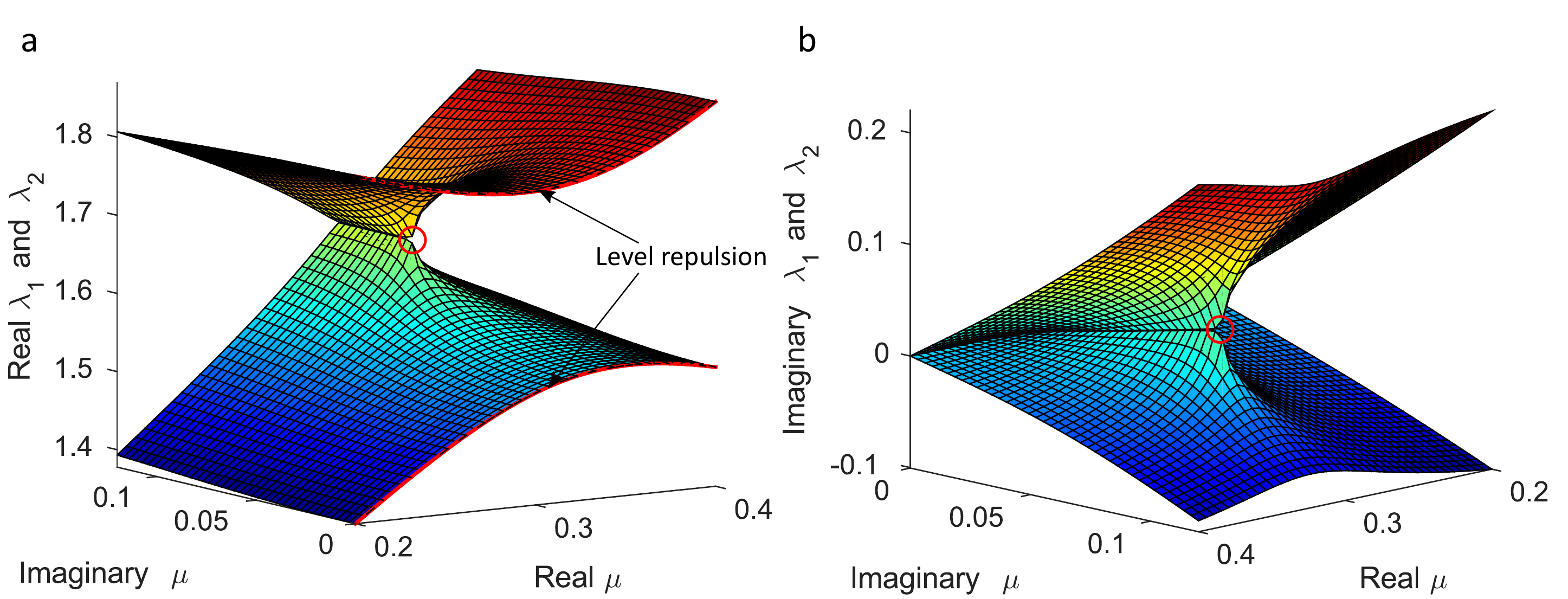}
\caption{(a),(b) Real and imaginary part of $\lambda_1$ and $\lambda_2$. Level repulsion on real $\mu$ axis is marked by the red curves. The real and imaginary part of the exceptional point are indicated by the red circles. The parameters in Eq. (\ref{egmatrix}): $\epsilon_1=1$, $\epsilon_1=2$, $\omega_1=2$, $\omega_1=-1$ and $\phi=\pi/25$.}\label{lambda1and2}
\end{figure}

As an example, if $\mathbf{A}$ is a $2\times 2$ matrix of the form 
%\cite{heiss1999phases}
\begin{equation}\label{egmatrix}
\mathbf{A}
=
\begin{bmatrix}
    \epsilon_1 & 0 \\
    0 & \epsilon_2 
\end{bmatrix}
+\mu \mathbf{U}
\begin{bmatrix}
    \omega_1 & 0 \\
    0 & \omega_2 
\end{bmatrix}
\mathbf{U}^\mathrm{T};\quad \mathbf{U}=
\begin{bmatrix}
    \cos\phi & -\sin\phi \\
    \sin\phi & \cos\phi 
\end{bmatrix}
\end{equation}
then it has an exceptional point at 
\begin{equation}
\mu_c=-\frac{\epsilon_1-\epsilon_2}{\omega_1-\omega_2}\exp(\pm 2\mathrm{i}\phi)
\end{equation}
$\mathbf{A}$ has two eigenvalues $\lambda_1,\lambda_2$ whose variations in the vicinity of the exceptional point are shown in Fig. (\ref{lambda1and2}). The exceptional point itself is marked with a red circle in the real and imaginary domains of the eigenvalues. The imaginary figure (Fig. \ref{lambda1and2}b) is flipped with respect to the imaginary $\mu$ axis for easy visualization of the exceptional point. Fig. (\ref{lambda1and2}a) shows both the coalescing of the two surfaces at the exceptional point and the associated phenomenon of level repulsion which follows on the real $\mu$ axis. The latter is emphasized by the red curve. On the real $\mu$ axis the complex parts of the eigenvalues are zero as expected.
Although Fig. (\ref{lambda1and2}) pertains to the behavior of the eigenvalues near the exceptional point of a $2\times 2$ matrix, it can be shown that a $2\times 2$ matrix captures the essential characteristics of an $N\times N$ matrix near an exceptional point if the exceptional point is isolated \cite{heiss1991transitional,heiss1999phases}. The discretized phononics eigenvalue matrix is such an $N\times N$ problem. To make a clear correspondence, note that phononic eigenvalue problem is self-adjoint for real values of the wavevector. The wavevector has the form $\mathbf{k}=0.5\mathbf{q}^1+Q_2\mathbf{q}^2$ where $0\leq Q_2\leq 0.5$ at all points along the $X-M$ direction in the bandstrucure in Fig. (\ref{unitcell}b). The eigenvalue matrix, along the $X-M$ direction, which includes the avoided cross point R, is, therefore, only dependent upon the parameter $Q_2$. In other words $\mathbf{\Lambda}^{(\mathbf{k})}$ is of the form $\mathbf{\Lambda}(Q_2)$ and is Hermitian (self-adjoint) as long as $Q_2$ is real (or follows the boundary of the IBZ). It has $N$ real eigenvalues $\lambda_n(Q_2)$ for a given value of real $Q_2$. However, $\mathbf{\Lambda}(Q_2)$ is non Hermitian for complex $Q_2$. We are interested in identifying the exceptional point which is giving rise to the avoided crossing at point R.
\begin{figure}[htp]
\centering
\includegraphics[scale=.5]{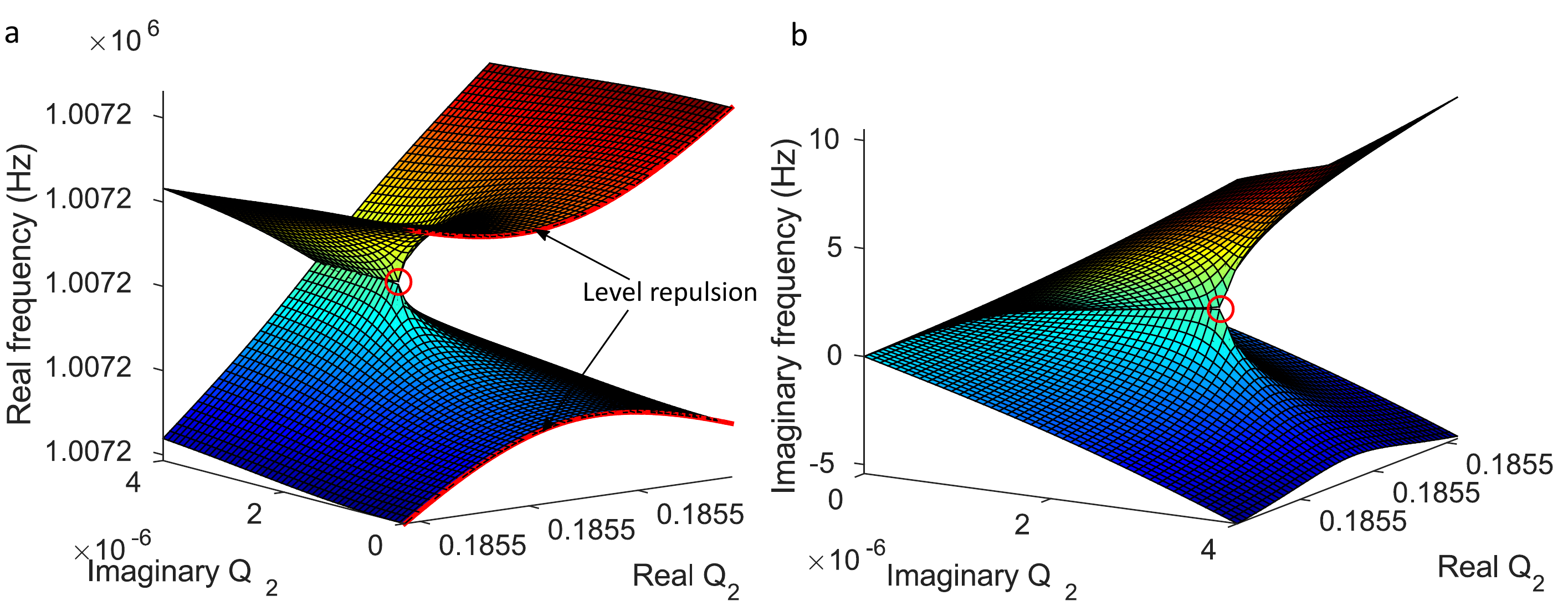}
\caption{(a),(b) Real and imaginary part of the frequencies near point R. The real and imaginary part of the exceptional point are marked by red circles.}\label{R1and2}
\end{figure}
Fig. (\ref{R1and2}) shows the behavior of the frequencies corresponding to the two avoided crossing branches when complex values of $Q_2$ are considered around point R. The required complex eigenvalues are calculated by using the generalized Rayleigh quotient\cite{lu2016variational}. As in Fig. (\ref{lambda1and2}), the imaginary $Q_2$ axis for the imaginary part of the frequency surfaces (Fig. \ref{R1and2}b) is flipped for the easy visualization of the coalescing point. The exceptional point is identified as $Q_2=0.1854865\pm \mathrm{i}2.6130959\mathrm\times 10^{-6}$. Even though Fig. (\ref{R1and2}) corresponds to the eigenvalues of a much larger matrix, its similarity with Fig. (\ref{lambda1and2}) shows that the behavior of the eigenvalues near the exceptional point is similar to those of a $2\times 2$ matrix. This similarity in appearance would clearly not have held if the larger matrix had another exceptional point in the vicinity\cite{heiss1991transitional}. 
\begin{figure}[htp]
\centering
\includegraphics[scale=.4]{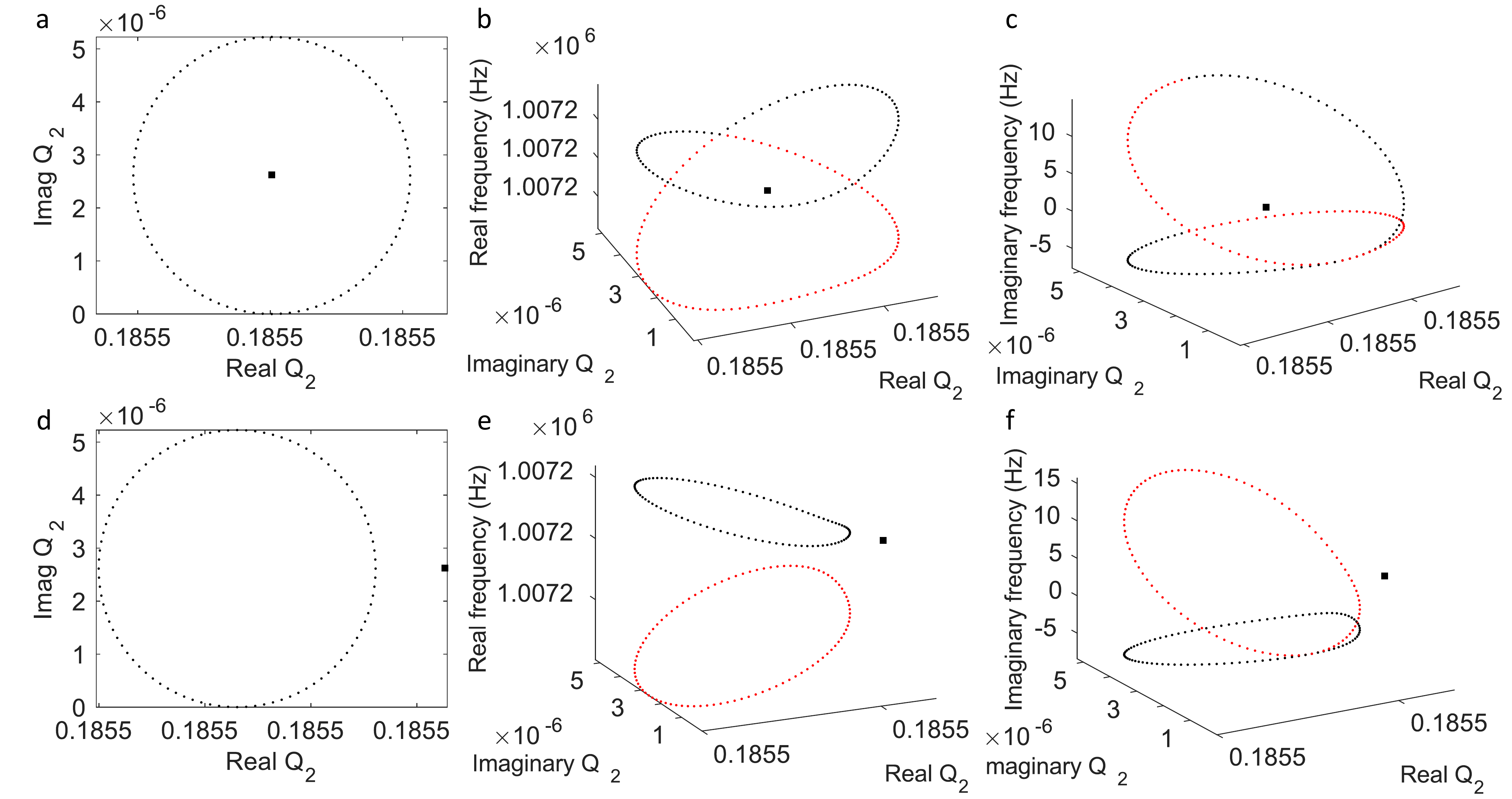}
\caption{(a) Closed curve in the wavenumber domain around the exceptional point. (b),(c) Real and imaginary parts respectively of the eigen-frequencies calculated along the curve in (a). (d) Closed curve in the wavenumber domain excluding the exceptional point. (e),(f) Corresponding real and imaginary parts of the eigen-frequencies calculated along the curve in (d). }\label{fcircle}
\end{figure}
The exceptional point forms a square root branch point\cite{heiss1990avoided,heiss1991transitional,heiss1999phases} and connects the two eigen-surfaces. A closed curve in the wavenumber domain around the exceptional point would require the two eigenfrequencies to switch if analytic continuation is to be maintained. Figs. (\ref{fcircle}b,c) show the movement of the real and imaginary parts of the two phononic eigenvalues as the wavenumber traces a closed curve around the exceptional point (Fig. \ref{fcircle}a). The exceptional point is indicated by the bold black dot whereas red and black dots represent the individual eigenvalues. There is a branch cut in Fig. (\ref{fcircle}a) which starts at the exceptional point and increases vertically along the positive imaginary $Q_2$ direction. The real parts of the eigenvalues are non-differentiable across this branch cut (Fig. \ref{fcircle}b) whereas the imaginary parts are discontinuous (Fig. \ref{fcircle}c). However, they can be analytically continued by switching the eigenvalues across the branch cut. This is evident from the smooth switching of the red curves into the black curves and vice-versa. If the closed curve in the wavenumber domain excludes the exceptional point (Fig. \ref{fcircle}d) then the corresponding eigenvalue curves remain separated and analytic (Figs. \ref{fcircle}e,f). 
\begin{figure}[htp]
\centering
\includegraphics[scale=.35]{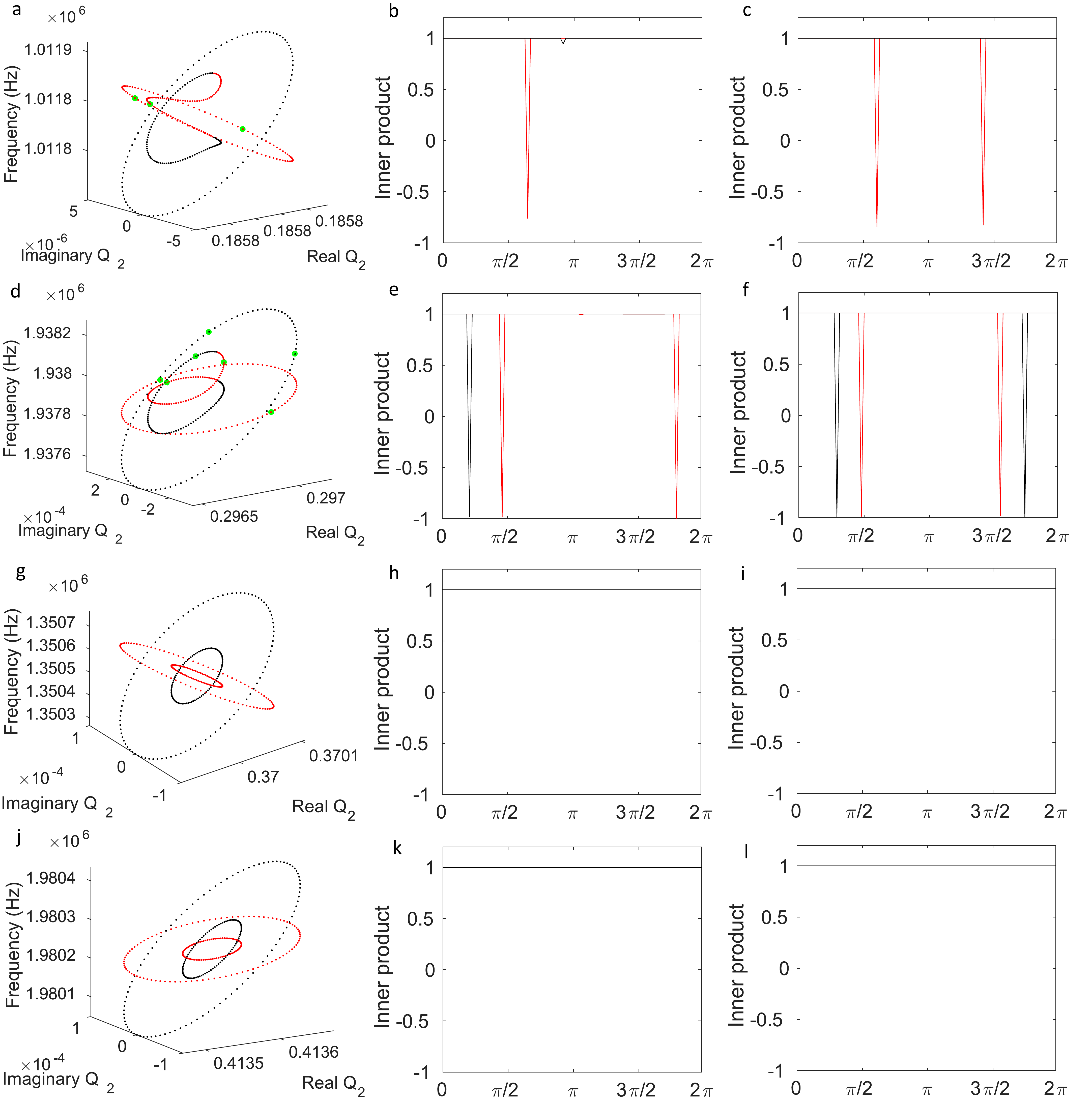}
\caption{(a,d,g,j): Eigenfrequency trajectories around points R, L$_1$, T, and L$_2$ in Fig. (\ref{unitcell}b) respectively. Eigenvectors change signs at locations which are indicated by bold green dots, (b, e, h, k): Inner products of adjacent eigenvectors over the smaller circles, (c,f,i,l): Inner products of adjacent eigenvectors over the larger circles.}\label{flipofsigns}
\end{figure}

In addition to the eigenvalue exchange phenomenon described above, the corresponding eigenvectors also behave in characteristic ways along closed curves which encircle an exceptional point \cite{heiss2000repulsion}. It is known, for instance, that for a simple $2\times 2$ matrix, one of the two (and only one) eigenvectors will flip sign around such a closed curve. For the more complex and higher-dimensional phononic case the situation is slightly different. To understand this difference, first consider Figs. (\ref{flipofsigns}a-c). Fig. (\ref{flipofsigns}a) shows the behavior of the two relevant phononic eigenvalues as two different closed curves are traversed in the vicinity of point R in Fig. (\ref{unitcell}b). The smaller curve encircles one exceptional point and the larger curve encircles the same exceptional point as well as its complex conjugate exceptional point.  Both curves are circles with the smaller circle centered at the exceptional point and the larger one centered on the real axis. We now calculate the eigenvector inner products over adjacent wavevector points as we traverse these circles. Fig. (\ref{flipofsigns}b) shows these inner products for the two eigenvectors over the smaller circle (as a function of the polar angle.) We note that while one of the two inner products (black curve) is very close to 1 over the entire circle, the other curve (red) achieves a value of -1 as we cross the branch cut. Over the closed curve, therefore, one of the eigenvectors has flipped its sign. Over the larger circle (Fig. \ref{flipofsigns}c) there are two sign flips for one of the eigenvectors essentially nullifying the flip. The location of these sign flips are also denoted by green dots in Fig. (\ref{flipofsigns}a). The situation is more complex for another exceptional point which exists at a higher frequency (Fig. \ref{flipofsigns}d corresponding to point L$_1$ in Fig. \ref{unitcell}b). In this case, even for the smaller circle (Fig. \ref{flipofsigns}e), there are multiple flips of sign for the two involved eigenvectors. However, even in this case the essential effect of the flips is to ensure that only one of the eigenvectors flips sign over the close curve. For the larger circle (Fig. \ref{flipofsigns}e), the essential effect of the sign flips is to ensure that over the closed curve there are no cumulative flips. Multiple flips, in these case, may be due to complex topology of the branch cut or other branch points (exceptional points) in the vicinity of the exceptional point considered. That there are more than one flips in the eigenvectors around an exceptional point does not necessarily defeat their use in our proposed band-sorting algorithm. What is important is that the flips do not happen at all when the closed curve does not encircle an exceptional point. This is shown in Figs. (\ref{flipofsigns}g-l). Figs. (\ref{flipofsigns}g,j) show four closed curves around two real cross points in the phononic bandstructure (T and L$_2$ respectively in Fig. \ref{unitcell}b). Figs. (\ref{flipofsigns}h,i,k,l) show the behavior of the inner products as these curves are traversed. There are no flips of sign as there are no exceptional points in the vicinity of real crosses. Numerical calculations show that the two eigenvalues in the case of point T coalesce on the real axis at $Q_1=0.37000376$, $Q_2=0$. For point L$_2$ they again coalesce on the real axis at $Q_1=Q_2=0.41390153$.

\section{Band Sorting for Phononic Crystals}

\begin{figure}[htp]
\centering
\includegraphics[scale=.3]{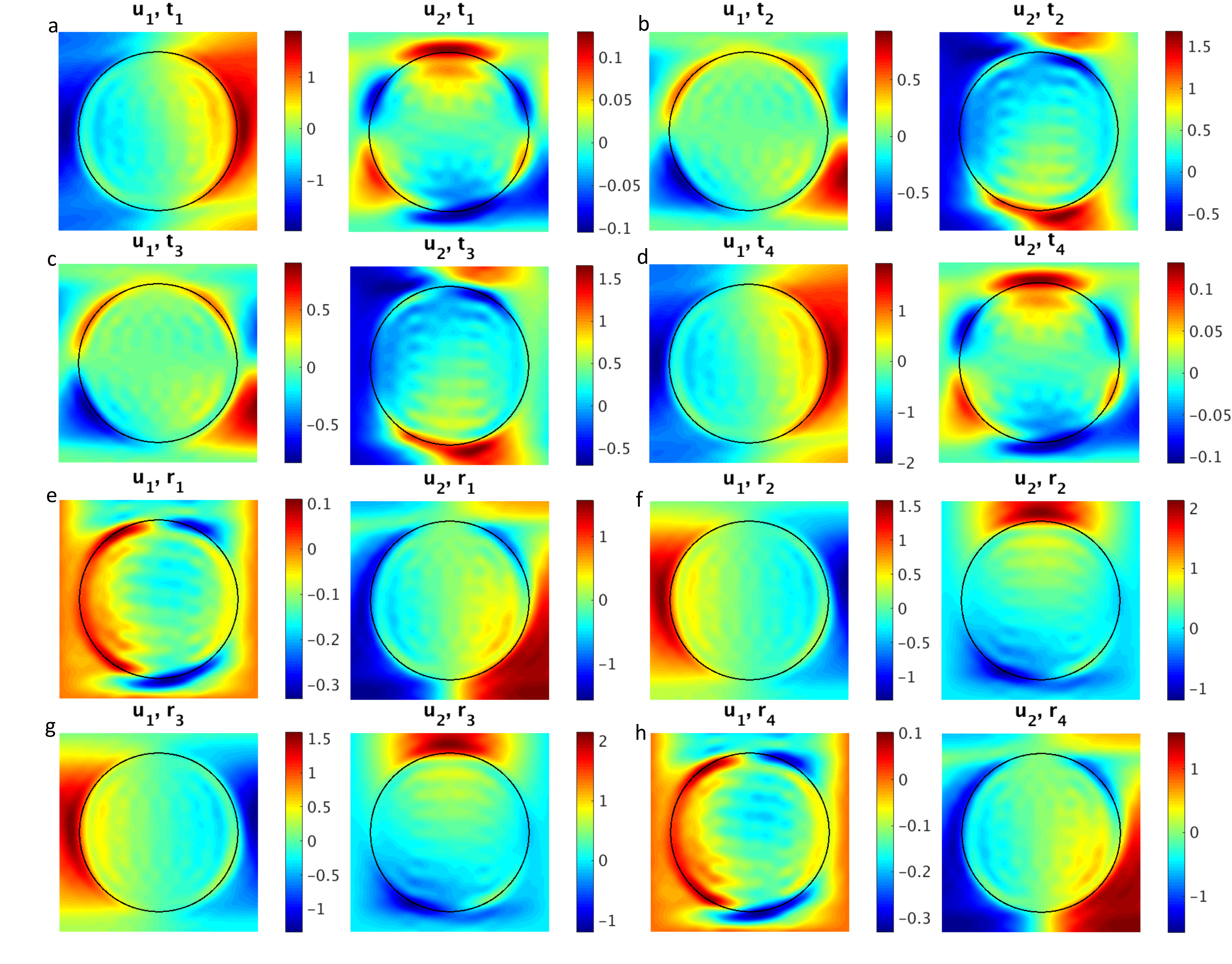}
\caption{(a),(b),(c),(d) Displacement fields in $x_1$ and $x_2$ direction at $t_1$-$t_4$. (e),(f),(g),(h) Displacement field in $x_1$ and $x_2$ direction at $r_1$-$r_4$.}\label{fmodeshape}
\end{figure}

Due to the presence of level repulsion, any band-sorting algorithm which depends solely upon eigenvector or modeshape continuation calculated on coarse wavenumber discretization is doomed from the start. This is because of the fact that in the vicinity of level repulsion not only do the eigenvalues avoid crossing each other but they also exchange their modeshapes. This ensures that any band-sorting algorithm which depends upon eigenvector continuation will mistake a level repulsion point as a real cross point if the modes of interest belong to two neighboring points on a rough grid of wavevectors. This phenomenon is explicitly shown in Fig. (\ref{fmodeshape}). Figs. (\ref{fmodeshape}a-d) are the displacement fields in the $x_1$ and $x_2$ directions respectively at points $t_1-t_4$ from Fig. (\ref{R1T1}a). The similarity of the modeshapes suggests that $t_4$ should be the continuation of $t_1$ and $t_3$ should be the continuation of $t_2$. This conclusion is correct since T is a real cross point. Figs. (\ref{fmodeshape}e-h) are the displacement fields in the $x_1$ and $x_2$ directions at points $r_1-r_4$ from Fig. (\ref{R1T1}b). In this case, the modeshape similarity suggests that $r_4$ should be the continuation of $r_1$ and $r_3$ should be the continuation of $r_2$. However, in this case the conclusion would be incorrect since point R is a level repulsion point as revealed by the fine discretization in Fig. (\ref{R1T1}d). Due to the exchange of eigenvectors, therefore, it is impossible to distinguish a level repulsion point from a real cross point purely from eigenvector orthogonality considerations.

\begin{figure}[htp]
\centering
\includegraphics[scale=.6]{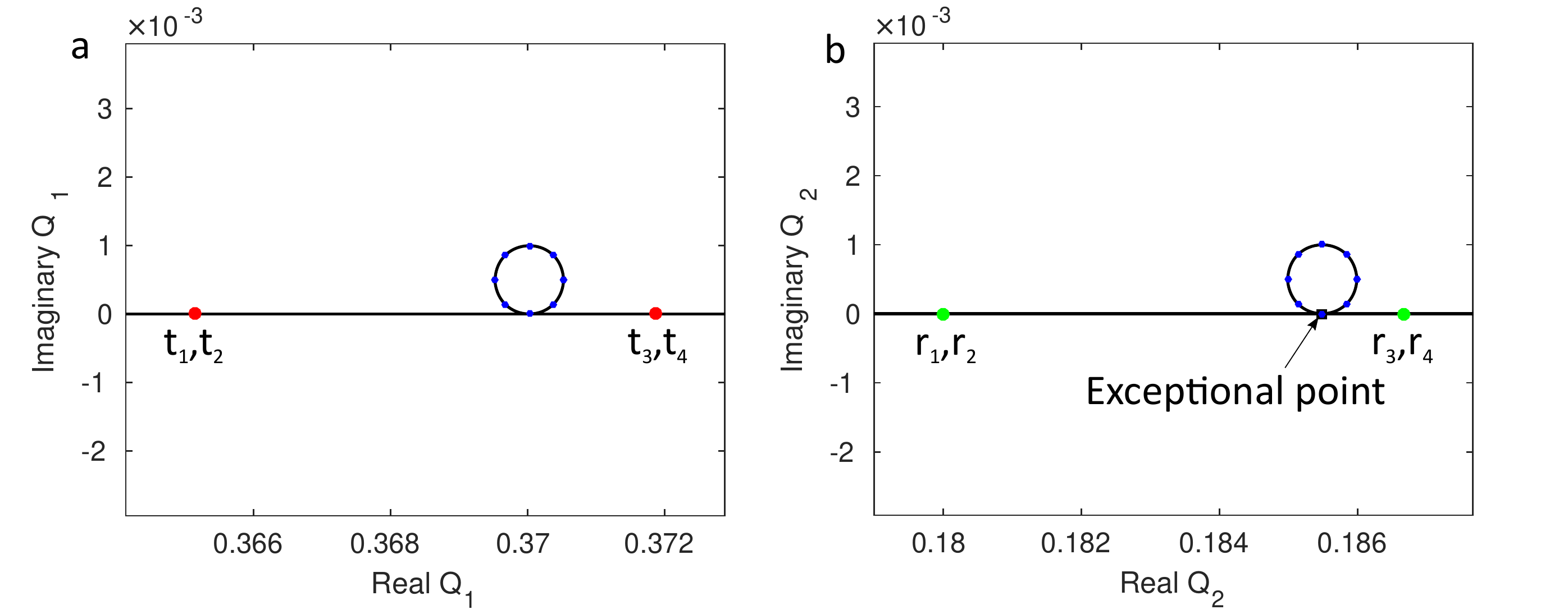}
\caption{Closed curves in the complex parameter domain near point T, (a), and  point R, (b). Blue dots mark the wavevetor points that are used in inner product calculations. The Exceptional point, marked by a small black square, is encircled by the curve in (b).}\label{circles}
\end{figure}

Here we propose a more efficient method of sorting the band-structure which forgoes the need for the zoom-in method. The idea rests on the observation from Fig. (\ref{flipofsigns}) that flips of signs in eigenvectors should be present as one considers the eigenvalue problem at discrete locations on a closed curve which encircles an exceptional point. No such flips would occur if the closed curve does not encircle an exceptional point. The process of implementation of this idea is further explained in Fig. (\ref{circles}) with respect to the two points discussed in Fig. (\ref{R1T1}). Figs. (\ref{circles}a,b) correspond to Figs. (\ref{R1T1}a,b) respectively. We consider the location of the hypothetical cross point (real or apparent) between two curves of interest and as determined by the coarse discretization. This is determined by finding the intersection of the two lines which approximate the curves in the region of interest\cite{heiss1990avoided,heiss1991transitional}. We draw a circle in the complex domain of the parameter. Its lower extremity lies at the hypothetical location of the cross-point and its radius is large enough to encompass the exceptional point if there is one in the vicinity. In general, the distance of the exceptional point from the real axis of the parameter is small and dependent upon the distance between the two curves in the region of repulsion\cite{heiss1990avoided}. In the case that level repulsion exists, such a circle will most likely encircle the exceptional point and exclude its complex conjugate exceptional point. The situation is shown in Fig. (\ref{circles}b) where the circle encircles the associated exceptional point. Fig. (\ref{circles}a) shows that the circle encircles no exceptional point since none exists in this case. The eigenvectors are now evaluated at some discrete points on the circle (marked in Fig. \ref{circles}a,b). Since the radius of the circle is still very small compared to the domain of the parameter (it is smaller than the coarse discretization), the set of two relevant eigenvectors at the discrete points will be nearly the same unless there is a flip of sign. In other words, the inner products of the eigenvectors corresponding to different points on the closed curve would be either very close to +1, 0, or -1. In the case of a real cross, all mutual inner products will be either close to +1 or 0. In the case of level repulsion there will be some inner products which have a value close to -1.

\begin{figure}[htp]
\centering
\includegraphics[scale=.55]{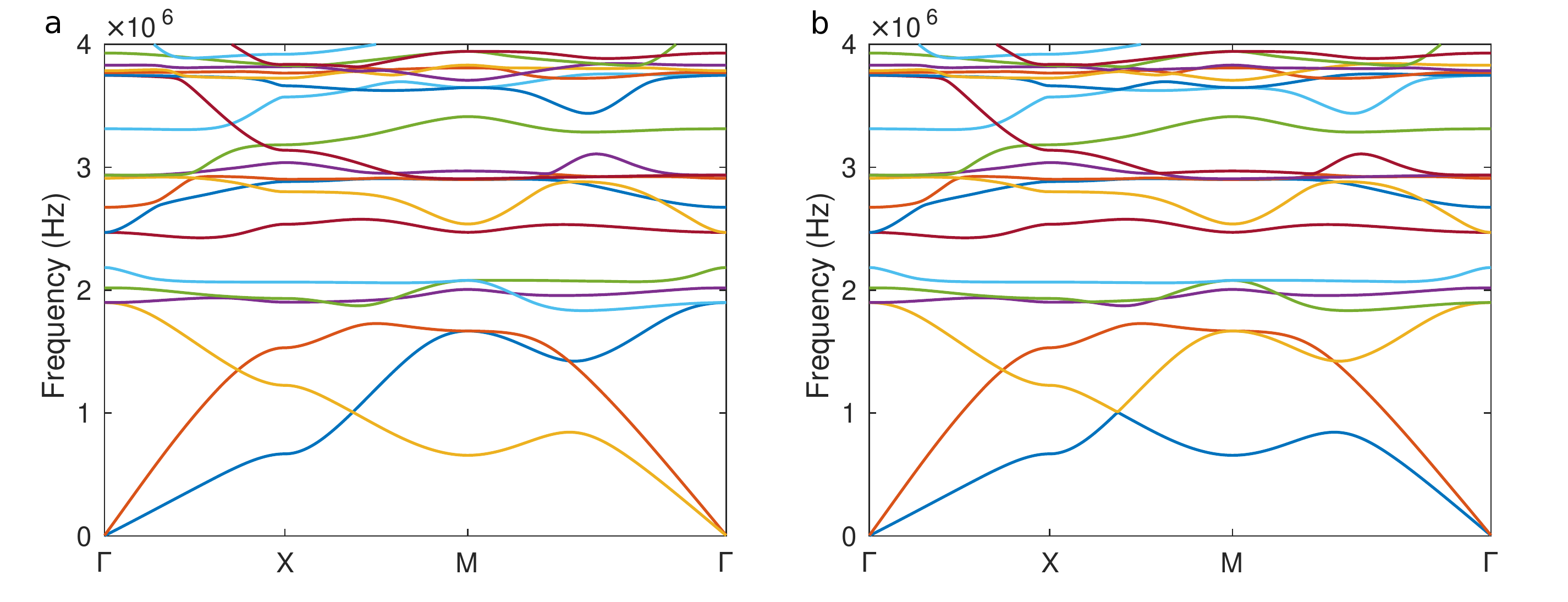}
\caption{(a) Sorted bands using eigenvector orthogonality on coarse wavenumber discretization. (b) Sorted bands by examining the existence of sign flip.}\label{bandsorting}
\end{figure}

To implement the above ideas into a band-sorting algorithm, the following processes are implemented: As a first step, we sort the eigenvalues in ascending order and rearrange the eigenvectors accordingly. Tentative mode continuation is determined by checking approximate orthogonality (\ref{eOrthoApp}), at adjacent points on the coarse discretization. This process results, for example, in the band-structure shown in Fig. (\ref{bandsorting}a) where individual colors indicate individual phononic bands. At this point the cross locations have been determined but level repulsion zones have not yet been distinguished from real crosses. We now form closed curves in the complex parametric domain near the vicinity of the crosses and analogous to the process shown in Fig. (\ref{circles}). By checking for flips of signs at discrete points over these closed curves and in accordance with the process described above, we distinguish real crosses from level repulsion zones. The results are shown in Fig. (\ref{bandsorting}b) where the level repulsion zones have been taken into account. Comparing it to Fig. (\ref{bandsorting}a), we can see that several mode continuations have been changed in the process. Band continuation of the first 6 curves agrees exactly with results in literature\cite{wu2004level}. For the above, the bandstrucuture is calculated by evaluating the eigenvalue problem at 256 wavevector points along the boundary of IBZ. To distinguish level repulsion zones from real crosses, circles of diameters $1\times10^{-3}$ are constructed in the complex domain of the parameter. Flips of signs are then checked at 9 discrete points along each circle (Fig. \ref{circles}).

\begin{figure}[htp]
\centering
\includegraphics[scale=.45]{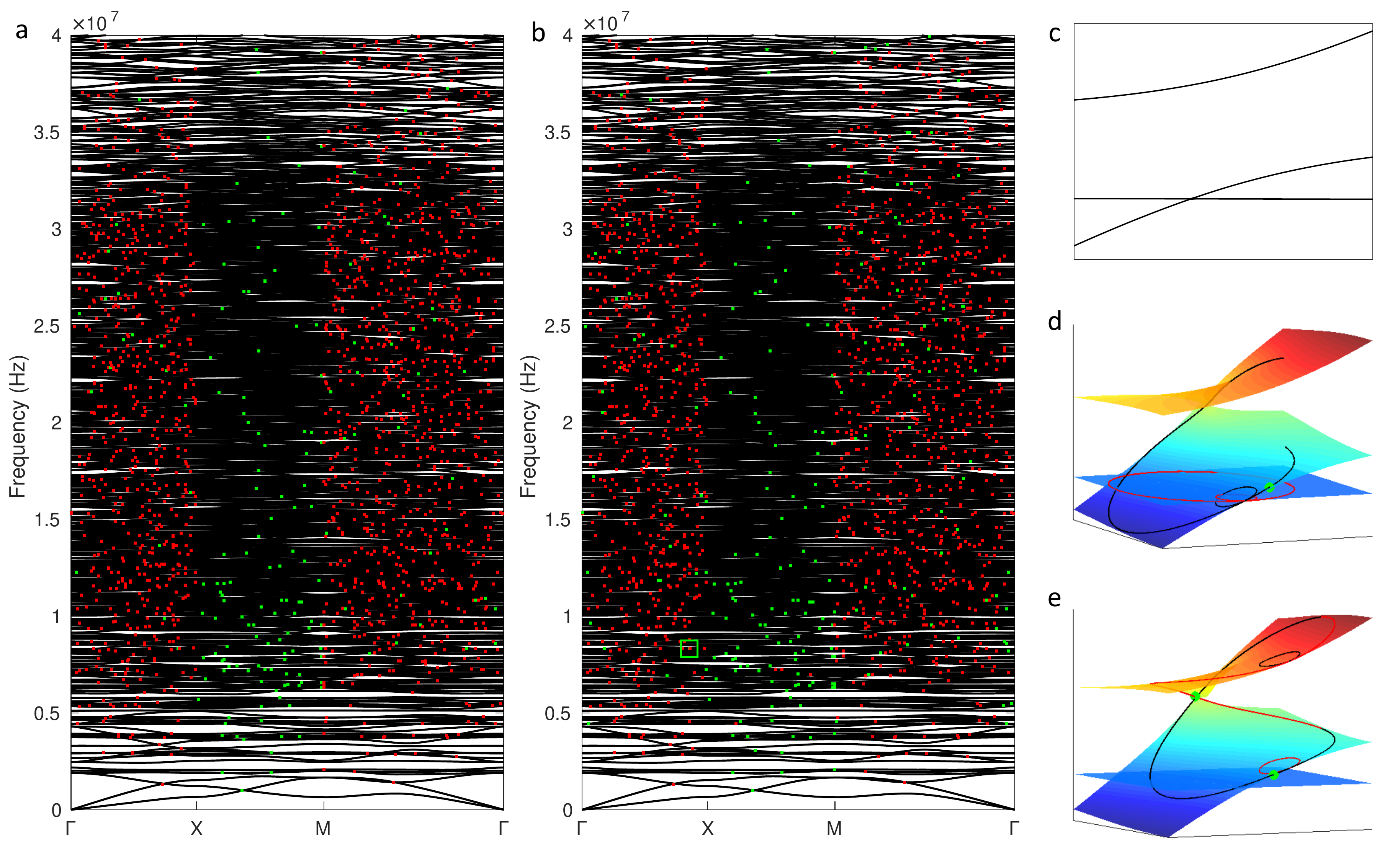}
\caption{(a) Sorted bandstructure using zoom-in method. (b) Sorted bandstructure by examining the existence of sign flip. Real cross points are marked by small red blocks and level repulsion points are marked by green blocks. (c) Zoom-in to the green square marked area in (b). Corresponding eigen-frequency surfaces of (c) in the complex wavevector domain and eigen-frequency solutions along circles near the real cross point (d) and the level repulsion (e).}\label{compare}
\end{figure}

In order to evaluate the overall performance of this method, we now present band-sorting results for larger number of phononic bands. The fully sorted bandstructutre for 683 phononic bands is shown in Fig. (\ref{compare}a). Sorting has been carried out using the zoom-in method here so there is no ambiguity in the results. There is a total of 1764 crosses (real or apparent) in the band-structure. Since the zoom-in method requires the evaluation of the eigenvalue problem at fine discretizations near the cross points, we had to evaluate the problem a total of 137592 times to achieve the sorting. Real cross points are marked by red markers and the level repulsion points are marked by green ones. It is interesting to note that all points between $X-M$, which would be determined as cross points using solely eigenvector continuity, turn out to be level repulsion points. The majority of points between $\Gamma-X$ and $M-\Gamma$ are real cross points. Fig. (\ref{compare}b) shows the sorted bandstructure based on examining the existence of sign flips. The latter process correctly identifies the crosses as real or apparent for 97.39\% of the total 1764 crosses. In effect, only 46 out of the 1764 total points are incorrectly identified by the flip of sign method. This essentially means, for the current problem, that 92.69\% of the 683 phononic bands have been assigned correct mode continuation through the flip of sign method. In contrast, if the bands were sorted in ascending order of magnitude, the accuracy of mode continuation would have been just 38.55\%. If the bands are sorted based only on the inner product continuity on the coarse grid then the accuracy would have been 79.08\%. The current implementation of the flip of sign method is roughly an order of magnitude faster than the zoom in method. Both the accuracy and the speed of the method can be improved as we discuss in the next section.

\section{Discussion and Conclusions}
In this paper we have considered the problem of level repulsion in the context of phononics and suggested a computationally efficient strategy to distinguish them from normal cross points. This process is essential for the correct sorting of the phononic bands and, subsequently, for the accurate determination of mode continuation, group velocities, and emergent properties which depend on them. We have shown that in the vicinity of the exceptional point the relevant phononic eigenvalue surfaces resemble the surfaces of a 2 by 2 matrix. Along a closed loop encircling the exceptional point the phononic eigenvalues are exchanged, just as they are for the 2 by 2 matrix case. However, the behavior of the associated eigenvectors is shown to be more complex in the phononic case. Along a closed loop around an exceptional point, we show that the eigenvectors can flip signs multiple times unlike a 2 by 2 matrix where the flip of sign occurs only once. Based upon this sign-flip property near an exceptional point, we have proposed a band-sorting technique for the accurate sorting of phononic eigenvalues. The method is roughly an order of magnitude faster than the fail-safe zoom-in method. As implemented, the method correctly identified $>$97\% of the total cross points in the example considered which translated into the correct mode continuation assignment for $>$90\% of the phononic branches.

There are several ways of further increasing both the accuracy and the speed of the sorting method. First, it should be noted that the only reason that the flip of sign method, as currently implemented, incorrectly identifies a level repulsion point as a real cross point is that the considered closed curve (circle in our case) fails to encircle an exceptional point. The circle misses encircling an exceptional point if the exceptional point is too close to the real axis and if the hypothetical location of the intersection of the two curves (which is also the lower extremity of the considered circle) is such that the exceptional points sits just below the circle boundary. This issue can potentially be remedied by considering other closed curves such as a semicircle instead of a circle. It may even be possible to forgo the closed curve completely and just evaluate the inner products at randomly chosen locations in the complex domain of the parameter and near the hypothetical cross point of the two curves under consideration. 

The current method never identifies a real cross point as a level repulsion point unless there is a real cross point in close vicinity of a level repulsion point. For example, Fig. (\ref{compare}c) shows that a real cross point interferes with level repulsion in the region marked by a small green square in Fig. (\ref{compare}b). The sorting algorithm draws two circles of the same diameter $1\times10^{-3}$ in the complex domain with one located at the real cross point and the other at the level repulsion point. These circles are large enough such that the one at the real cross point ends up encircling the exceptional point which belongs to the level repulsion zone. Figs. (\ref{compare}d,e)show that both eigenvalue trajectories along the two curves pass the branch cut where the eigenvectors flip signs. Therefore, this real cross point is mistakenly identified as a level repulsion by the algorithm. Such a situation can be remedied by employing a smaller closed curve. However, it must be noted that a smaller curve will, in general, increase the chance that it may fail to encircle an exceptional point when it exists. Therefore, further strategies should take into account this situation. Finally, we have chosen to evaluate 9 discrete points along the closed curves to check for sign-flips. It is entirely possible that one can implement an algorithm which maintains the accuracy while evaluating the eigenvalue problem at far fewer locations. This would further speed-up the process of band-sorting.

\section*{Acknowledgment}
A.S. acknowledges support from the NSF CAREER grant \# 1554033 to the Illinois Institute of Technology.

%\bibliography{ReferencesBib}

\end{document}